\documentclass{aa}

\usepackage{graphicx}
\usepackage{txfonts}
\usepackage{natbib}
\usepackage{longtable}
\bibpunct{(}{)}{;}{a}{}{,}
\begin{document}

\title{Relativistic beaming and gamma-ray brightness of blazars}

  \author{T.~Savolainen\inst{1} \and D.~C.~Homan\inst{2} \and
    T.~Hovatta\inst{3,4} \and M.~Kadler\inst{5,6,7} \and
    Y.~Y.~Kovalev\inst{8,1} \and M.~L.~Lister\inst{3} \and E.~Ros\inst{1,9}
    \and J.~A.~Zensus\inst{1}}

  \offprints{T.~Savolainen}

  \institute{Max-Planck-Institut f\"ur Radioastronomie, Auf dem
             H\"ugel 69, D-53121, Bonn, Germany\\
             \email{tsavolainen@mpifr-bonn.mpg.de} 
             \and
             Department of Physics and Astronomy, Denison University,
             Granville, OH 43023, USA 
             \and
             Department of Physics, Purdue University, 525 Northwestern
             Avenue, West Lafayette, IN 47907, USA
             \and
             Mets\"ahovi Radio Observatory, Helsinki University of
             Technology TKK, Mets\"ahovintie 114, 02540 Kylm\"al\"a, Finland
             \and
             Dr. Karl Remeis-Observatory \& ECAP, Friedrich-Alexander
             University Erlangen-Nuremberg, Sternwartstr. 7, 96049 Bamberg,
             Germany
             \and
             CRESST/NASA Goddard Space Flight Center, Greenbelt, MD 20771, USA
             \and
             Universities Space Research Association, 10211 Wincopin Circle,
             Suite 500 Columbia, MD 21044, USA 
             \and
             Astro Space Center of Lebedev Physical Institute, Profsoyuznaya
             84/32, 117997 Moscow, Russia
             \and
             Departament d'Astronomia i Astrof\'isica, Universitat de
             Val\`encia, E-46100 Burjassot, Valencia, Spain} 

\date{Received $<$date$>$; accepted $<$date$>$}

\abstract{}{We investigate the dependence of $\gamma$-ray brightness of
  blazars on intrinsic properties of their parsec-scale radio jets and the
  implication for relativistic beaming.}{By combining apparent jet speeds
  derived from high-resolution VLBA images from the MOJAVE program with
  millimetre-wavelength flux density monitoring data from Mets\"ahovi Radio
  Observatory, we estimate the jet Doppler factors, Lorentz factors, and
  viewing angles for a sample of 62 blazars. We study the trends in these
  quantities between the sources which were detected in $\gamma$-rays by the
  $Fermi$ Large Area Telescope (LAT) during its first three months of science
  operations and those which were not detected.}{The LAT-detected blazars have
  on average higher Doppler factors than non-LAT-detected blazars, as has been
  implied indirectly in several earlier studies. We find statistically
  significant differences in the viewing angle distributions between
  $\gamma$-ray bright and weak sources. Most interestingly, $\gamma$-ray
  bright blazars have a distribution of comoving frame viewing angles that is
  significantly narrower than that of $\gamma$-ray weak blazars and centred
  roughly perpendicular to the jet axis. The lack of $\gamma$-ray bright
  blazars at large comoving frame viewing angles can be explained by
  relativistic beaming of $\gamma$-rays, while the apparent lack of
  $\gamma$-ray bright blazars at small comoving frame viewing angles, if
  confirmed with larger samples, may suggest an intrinsic anisotropy or
  Lorentz factor dependence of the $\gamma$-ray emission.}{}


  \titlerunning{}
 
  \maketitle

%

\section{Introduction}
One of the most important discoveries of the Energetic Gamma-Ray Experiment
Telescope (EGRET) on-board the {\it Compton Gamma-Ray Observatory} in the
1990s was the detection of over 65 active galactic nuclei (AGN) at photon
energies above 100\,MeV \citep{har99,mat01}. The detected sources were almost
exclusively blazars, a class of highly variable AGN comprised of flat spectrum
radio quasars and BL Lac objects. The distinctive characteristic of blazars is
a relativistic jet oriented close to our line-of-sight. Synchrotron radiation
of energetic electrons in the jet dominates the low energy end of the blazar
spectral energy distribution. This emission is strongly beamed due to
relativistic effects which increase the observed flux density of a stationary
jet by a factor of $\delta^{2-\alpha}$ and that of distinct ``blobs'' in the
jet by a factor of $\delta^{3-\alpha}$.  Here $\delta$ is the jet Doppler
factor and $\alpha$ is the spectral index defined as $S_\nu \propto
\nu^{+\alpha}$ \citep{bla79}. The Doppler factor is defined as $\delta =
[\Gamma (1-\beta \cos \theta)]^{-1}$, where $\Gamma = (1-\beta^2)^{-1/2}$ is
the bulk Lorentz factor, $\beta$ is the jet speed divided by the speed of
light, and $\theta$ is the angle between the jet and our line-of-sight. The
requirement that the $\gamma$-ray-bright sources are transparent to
$\gamma\gamma$ pair production, together with their small sizes deduced from
the fast $\gamma$-ray variability, strongly suggest that the $\gamma$-ray
emission originates in the jet and is also relativistically beamed, in the
same manner as the radio emission \citep{mon95,mat93}. The many correlations
found between the $\gamma$-ray emission detected by EGRET and the
radio/mm-wave properties of blazars further support this scenario
\citep{val95, jor01b, jor01a, lah03, kel04, kov05}. Since the $\gamma$-ray
spectrum is typically steeper than the radio spectrum \citep{abd09b}, it is
possible that the $\gamma$-ray emission is even more enhanced by Doppler
boosting than the radio emission.

Although inverse Compton (IC) scattering of soft photons off relativistic
electrons in the jet is currently the favoured model for the $\gamma$-ray
emission, there is a substantial controversy within this model about the
origin of the target photon field and the location of the emission site. The
seed photons could be, for example, synchrotron photons emitted by the same
electrons which scatter them later, \citep[synchrotron self-Compton model,
SSC;][]{mar92, blo96} or synchrotron photons emitted by electrons in a
different layer of the jet \citep{ghi05}. The seed photons can also originate
in sources external to the jet like the accretion disk, the broad line region
clouds or the dust torus \citep[external Compton
model;][]{der92,ghi96,sik94,bla00}. Beside these leptonic models there are
also a number of models where $\gamma$-rays are produced by hadronic processes
initiated by relativistic protons co-accelerated with electrons
\citep[e.g.,][]{man93}.

The Large Area Telescope (LAT) on-board the {\it Fermi Gamma-ray Space
  Telescope} is a successor to EGRET, with much better sensitivity, larger
energy range (up to 300 GeV), better angular resolution, and larger
field-of-view \citep{atw09}. In only its first three months of science
operations, the LAT detected 205 bright $\gamma$-ray sources at $>10\sigma$
level \citep{abd09a}, 116 of which are associated with AGN at high galactic
latitudes ($|b| \ge 10^\circ$) \citep{abd09b}. Here we refer to these 116
sources as ``LAT-detected sources''. Based on the long-term monitoring of the
radio jet motions with the Very Long Baseline Array (VLBA), it was recently
shown that the LAT-detected quasars have significantly faster apparent jet
speeds and higher core brightness temperatures than non-LAT-detected quasars
\citep{lis09b,kov09}. The LAT $\gamma$-ray photon flux also correlates with
the compact radio flux density and the flares in $\gamma$-rays and radio seem
to happen in the VLBI cores within a typical apparent time separation of up to
several months \citep{kov09}. The $\gamma$-ray bright blazars also have
larger-than-average apparent jet opening angles \citep{pus09}. These findings
indicate that $\gamma$-ray bright blazars are likely more Doppler boosted than
$\gamma$-ray faint ones, but what remains unknown is a possible dependence of
{\it intrinsic} $\gamma$-ray luminosity on parsec-scale jet properties such as
the bulk Lorentz factor or the viewing angle in the comoving frame of the jet
\citep{lis09b}. In this paper we confirm the connection between the
$\gamma$-ray brightness and the Doppler factor of the parsec-scale jet for a
sample of 62 blazars.  We have also combined measurements of the apparent jet
speeds and temporal variability at mm-wavelengths to derive jet Lorentz factor
and the viewing angle in the observer's frame and in the frame comoving with
the jet for 57 blazars in order to study how the $\gamma$-ray detection
probability depends on these intrinsic jet parameters.

\begin{figure*}
\centering
\includegraphics[width=0.75\textwidth]{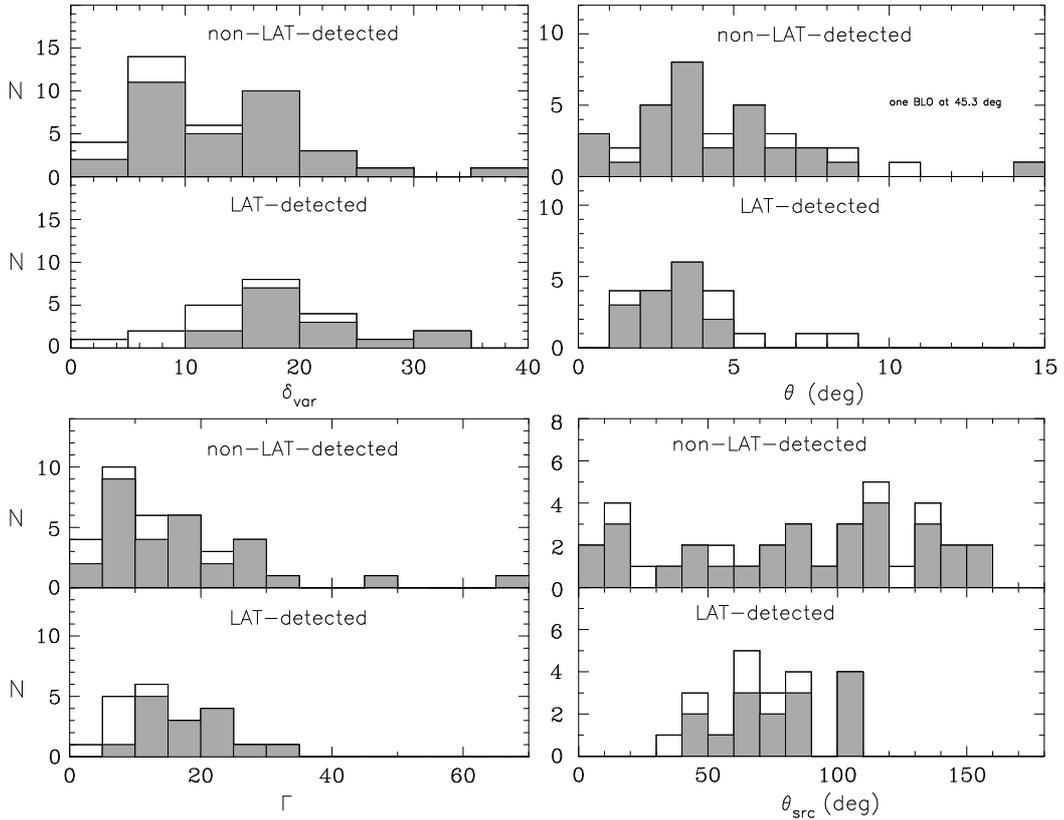}
\caption{{\it Upper left:} Variability Doppler factor ($\delta_\mathrm{var}$)
  distributions for non-LAT-detected (upper sub-panel) and LAT-detected (lower
  sub-panel) blazars in the Mets\"ahovi-MOJAVE sample. Quasars are denoted by
  shaded bins. {\it Lower left:} Bulk Lorentz factor ($\Gamma$)
  distributions. {\it Upper right:} Distributions of the angles between the
  jet and our line-of-sight in the observer's frame ($\theta$). {\it Lower
    right:} Distributions of the viewing angles in the comoving frame of the
  jet ($\theta_\mathrm{src}$). The values were calculated assuming
  $\Lambda$CDM cosmology with $H_0 = 71$\,km\,s$^{-1}$\,Mpc$^{-1}$,
  $\Omega_\mathrm{M} = 0.27$, and $\Omega_\mathrm{\Lambda} = 0.73$.}
\label{fig1}
\end{figure*}

\section{Sample definition and data}
The MOJAVE program, together with its predecessor, the VLBA 2\,cm Survey, has
continuously monitored the structural changes in the parsec-scale jets of the
compact extragalactic radio sources with the VLBA since 1994
\citep{kel04,lis09a}. The program provides accurate measurements of the
apparent jet speeds, which are typically superluminal \citep{lis09c}. The
current monitoring list includes a statistically complete,
flux-density-limited sample of all 135 sources that lie above J2000
declination $-20^\circ$ and had a 15\,GHz compact flux density larger than
1.5\,Jy (2\,Jy for sources below declination of $0^\circ$) at any time between
1994.0 and 2004.0. Since the sample is selected based on the relativistically
beamed, compact jet emission, it mostly comprises blazars and thus can be
compared with the sample of $\gamma$-ray-loud AGN.

The Mets\"ahovi Radio Observatory in Finland has regularly monitored the flux
density variability of bright northern hemisphere blazars at 22 and 37\,GHz
since the early 1980s, and decade or longer flux density curves exist for
$\sim100$ sources, which roughly approximate a flux-density-limited sample at
22\,GHz \citep{ter04}. From these data \citet{hov09} were able to estimate the
Doppler beaming factors of 87 sources by applying a light-travel time
argument. If one assumes that the variability timescale of the mm-wave flares
seen in these sources corresponds to the light-travel time across the emission
region \citep{jor05}, and that the intrinsic brightness temperature of
the source is limited to the equipartition value, $T_\mathrm{b,int} \lesssim 5
\times 10^{10}$\,K \citep{rea94}, measuring the flare timescale and amplitude
provides a way to estimate the variability Doppler factor,
$\delta_\mathrm{var}$. By combining equations 2 and 3 in \citet{hov09} one
gets
\begin{equation}
  \delta_\mathrm{var} = \Big( 1.47 \times 10^{13} \frac{\Delta S_\mathrm{max}
    d_\mathrm{L}^2}{\nu^2 \Delta t^2 (1+z)  T_\mathrm{b,int}} \Big) ^{1/3},
\label{dvar}
\end{equation}
where $\Delta S_\mathrm{max}$ is the flare amplitude in Janskys, $\nu$ is the
observing frequency in GHz, $\Delta t$ is the timescale of the fastest flare
in the source in days, $d_\mathrm{L}$ is the luminosity distance in Mpc, and
$z$ is the redshift. This method of estimating $\delta_\mathrm{var}$ relies on
the assumption that every source occasionally reaches $T_\mathrm{b,int} = 5
\times 10^{10}$\,K, that the Doppler factor is constant with time, and that
the flux density curve at mm-wavelengths is a superposition of flares with
mixed duration and amplitude \citep{hov09}. Thus, identifying the sharpest,
fastest flare over a long period of time gives the best estimate of the true
Doppler factor. We note that there is an inherent limit to the maximum
observable $\delta_\mathrm{var}$ of about 40--50 due to the time sampling of
Mets\"ahovi flux curves. We have carried out Monte Carlo simulations, however,
which show that this limit does not affect the statistical results reported in
this paper.

Of the 87 sources in \citet{hov09}, 60 are blazars that belong to the
flux-density-limited MOJAVE sample, have reliable redshifts, and are at
galactic latitude $|b| \ge 10^\circ$. In this paper, we refer to them as the
Mets\"ahovi-MOJAVE (MM) sample. Two additional sources, the BL Lac object
\object{B0109+224} and the quasar \object{B1334$-$127}, which were not
originally listed in \citet{hov09}, were also included in the sample since new
data allowed determination of their $\delta_\mathrm{var}$. With these
additions, the sample contains 48 quasars and 14 BL Lac objects.

There are 23 (37\%) sources in the MM sample that are associated with bright
LAT-detected $\gamma$-ray sources \citep{abd09b}: 15 quasars (31\% of the
quasars in the MM sample) and 8 BL Lacs (57\% of the BL Lacs). The fraction of
LAT-detected sources in the MM sample (37\%) is slightly higher than in the
full flux-density-limited MOJAVE sample (24\%) \citep{lis09b}, but the
difference is not statistically significant. An Anderson-Darling (A--D)
variant of the Kolmogorov-Smirnov test \citep{pre92} does not show a
significant difference between the redshift distributions of LAT-detected and
non-LAT-detected sources in the MM sample. For all the statistical tests
reported in this paper, we adopt a significance level of 0.05 for rejecting
the null hypothesis. We use the non-parametric A--D test throughout as a test
for similarity of two observed distributions, since it is more sensitive to
differences in the distribution tails than the original Kolmogorov--Smirnov
test.

\section{Results}

\subsection{Doppler factor and  Lorentz factor distributions}
The distributions of the variability Doppler factor for the non-LAT-detected
and the LAT-detected blazars in the MM sample are shown in the upper left
panel of Fig.~\ref{fig1}, and the numerical data are listed in
Table~\ref{tab1}. The LAT-detected sources have on average significantly
higher $\delta_\mathrm{var}$ values than non-LAT-detected sources: the mean
$\delta_\mathrm{var}$ is $17.1\pm1.6$ for the former and $12.2\pm1.2$ for the
latter.  For quasars only, the mean $\delta_\mathrm{var}$ are $19.8\pm1.7$ for
LAT-detected and $13.4\pm1.3$ for non-LAT-detected, and there are no
LAT-detected quasars with $\delta_\mathrm{var} < 10$. A one-sided Student's
t-test confirms that the LAT-detected sources have a higher mean
$\delta_\mathrm{var}$ than the non-LAT-detected ones with the probability of
99.3\% for blazars, and 99.7\% for quasars only.

\addtocounter{table}{1}

Since the apparent jet speed ($\beta_\mathrm{app}$) and the Doppler factor
both depend on the bulk Lorentz factor and the viewing angle of the jet, it is
possible to calculate the latter quantities if the measured
$\delta_\mathrm{var}$ and $\beta_\mathrm{app}$ both correspond to the same
underlying flow speed \citep{hov09}:
\begin{equation}
\Gamma = \frac{\beta_\mathrm{app}^2 + \delta_\mathrm{var}^2+1}{2\delta_\mathrm{v
ar}}, \quad \theta = \arctan \Big( \frac{2\beta_\mathrm{app}}{\beta_\mathrm{app}^2
  + \delta_\mathrm{var}^2 -1} \Big).
\label{gamma}
\end{equation}
Using the data gathered in the MOJAVE program, we have measured
$\beta_\mathrm{app}$ for 57 out of 62 sources in the MM sample by tracking the
motion of the bright features in the jet \citep{lis09c}. As shown by
\citet{lis09c}, there is a characteristic flow speed for each jet that can be
traced by $\beta_\mathrm{app}$. We consider the fastest, non-accelerating,
radially moving component to be the most representative of the speed of the
underlying flow close to the jet core, where the mm-wavelength outbursts take
place \citep{sav02}. The speeds are listed in Table~\ref{tab1} together with
the calculated jet properties. The Lorentz factor distributions for
non-LAT-detected and LAT-detected blazars are shown in the lower left panel of
Fig.~\ref{fig1}. Although there seems to be a lack of LAT-detected quasars
with small Lorentz factors (there are no LAT-detected quasars with $\Gamma <
8$), an A--D test does not detect a statistically significant difference
between the distributions -- neither for blazars nor for quasars alone.

\subsection{Viewing angle distributions}
The (observer's frame) viewing angle distributions are plotted in the upper
right panel of Fig.~\ref{fig1}. The $\gamma$-ray bright sources seem to have a
narrower $\theta$-distribution than $\gamma$-ray weak sources. The
LAT-detected quasars all have $1^\circ < \theta < 5^\circ$, while a
significantly wider range of values is covered for the non-LAT-detected
quasars: the observed $\theta$-distributions of LAT-detected and
non-LAT-detected quasars have a probability of $p=0.04$ of being drawn from
the same parent distribution, according to an A--D test. For the whole blazar
sample, the difference is not statistically significant. The mean value of
$\theta$ is slightly smaller for the LAT-detected quasars ($2.9\pm0.3^\circ$)
than for the non-LAT-detected ones ($4.4\pm0.6^\circ$), but this difference is
not statistically significant ($p=0.06$ according to a Student's t-test).

\begin{figure*}
\centering
\includegraphics[width=0.49\textwidth]{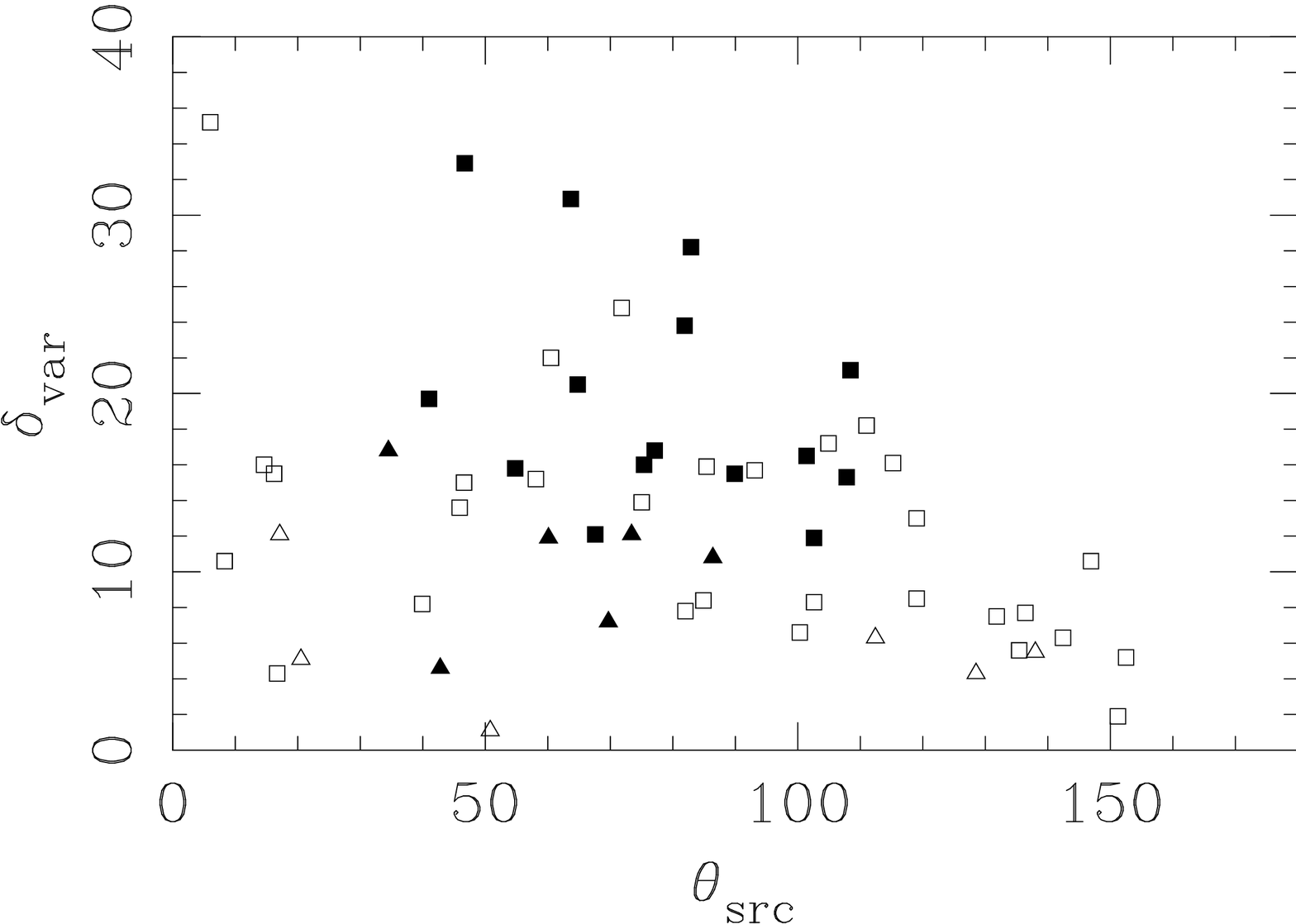}
\includegraphics[width=0.49\textwidth]{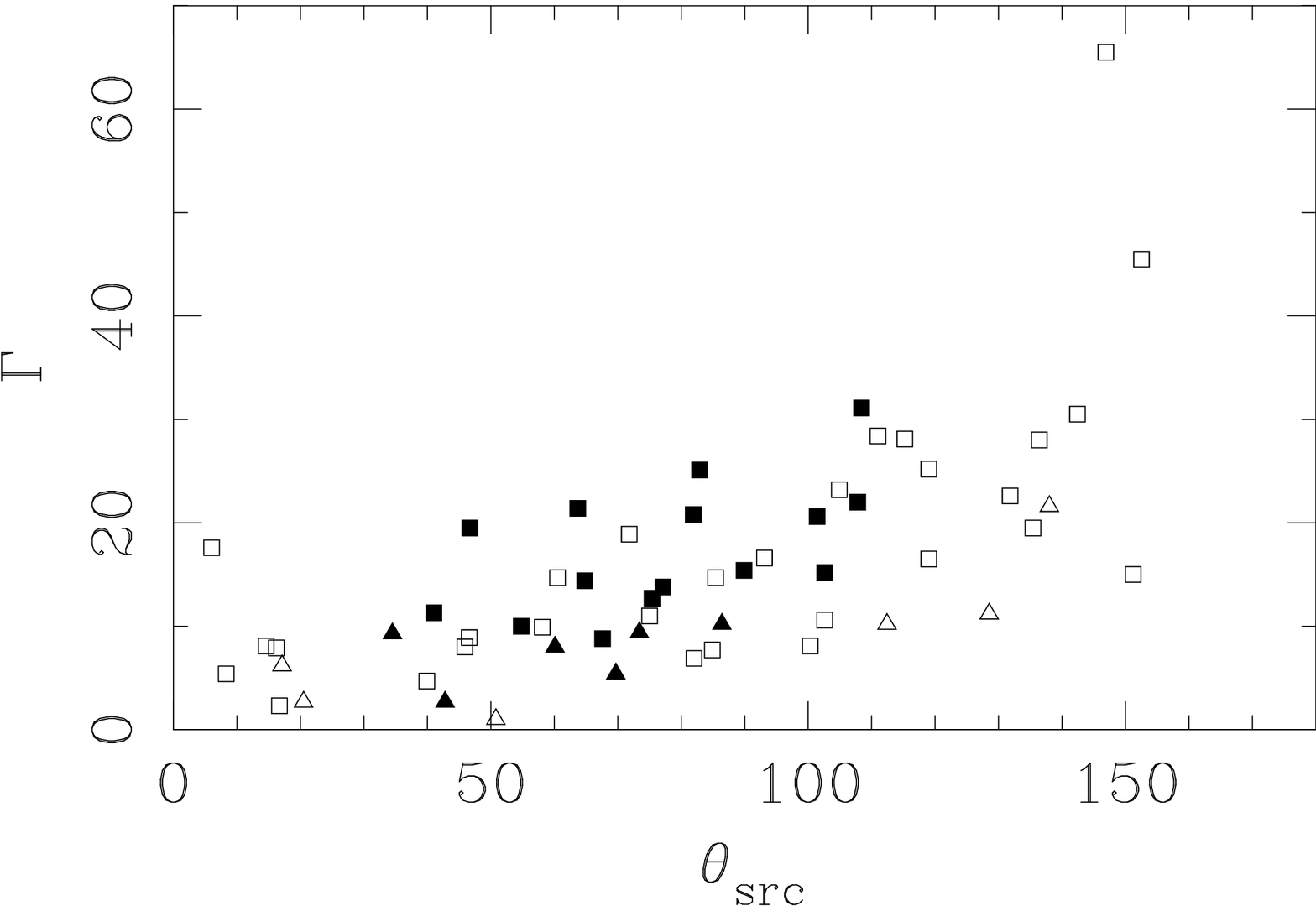}
\caption{{\it Left:} Variability Doppler factor as a function of a comoving
  frame viewing angle (in degrees) in the MM sample. Squares and triangles
  denote quasars and BL Lacs, respectively. Filled symbols are LAT-detected
  sources. {\it Right:} Jet Lorentz factor as a function of a comoving frame
  viewing angle in the MM sample.}
\label{fig2}
\end{figure*}

Because of relativistic aberration, the photons arriving to us at an angle
$\theta$ with respect to the jet flow direction were emitted from the jet at
an angle $\theta_\mathrm{src}$ in the {\it frame comoving with the jet}:
\begin{equation}
\theta_\mathrm{src} = \arccos \Big( \frac{\cos \theta - \beta}{1-\beta \cos
  \theta} \Big).
\label{aber}
\end{equation}
The distributions of viewing angles in the comoving frame are shown in the
lower right panel of Fig.~\ref{fig1}, where we see that the
$\theta_\mathrm{src}$-distribution is much narrower for the LAT-detected
blazars than for the non-LAT-detected blazars. The comoving fluid frame
viewing angles of $\gamma$-ray bright blazars are confined to a range between
$30^\circ$ and $110^\circ$ with a median value of $\sim 75^\circ$, while the
$\theta_\mathrm{src}$-distribution of $\gamma$-ray weak blazars ranges from
$0^\circ$ to $160^\circ$. Both small and large values of $\theta_\mathrm{src}$
are missing from the LAT-detected sample. An A--D test gives a probability of
$p=0.02$ that both samples are derived from the same parent distribution and
rejects the null hypothesis. For the smaller sample of quasars only, an A--D
test gives $p=0.09$ and does not reject the null hypothesis.

\section{Discussion} 

The variability Doppler factors from \citet{hov09} were determined over a
multi-year monitoring program while the $\gamma$-ray detections used in our
analysis are based on just three-months of LAT monitoring.  The sharp
distinction seen in the $\delta_\mathrm{var}$ values between the LAT-detected
and non-LAT-detected sources strongly supports the idea that
$\delta_\mathrm{var}$ must remain fairly constant with time. Since the
LAT-detected and non-LAT-detected sources are treated in the exact same way
with respect to the derivation of their $\delta_\mathrm{var}$, any temporal
variation in the Doppler factor (or in $T_\mathrm{b,int}$) would thus only
serve to destroy the possible correlation and could not create one. Therefore,
we consider the result regarding higher $\delta_\mathrm{var}$ for the
LAT-detected blazars to be very robust.

A similar result regarding the high $\delta_\mathrm{var}$ of $\gamma$-ray
bright sources was found earlier using the less uniform EGRET data
\citep{lah03}. The simplest and arguably most likely interpretation is that
the $\gamma$-ray bright blazars are indeed systematically more Doppler boosted
than the $\gamma$-ray weak ones. This interpretation is compatible with the
LAT-detected quasars having faster apparent jet speeds \citep{lis09b}, wider
apparent jet opening angles \citep{pus09}, and higher VLBI brightness
temperatures \citep{kov09}. An alternative interpretation would be that, for
some reason, the intrinsic brightness temperature is systematically about a
factor of three higher in the LAT-detected blazars than in the
non-LAT-detected ones. In the latter case our assumption of a constant
limiting brightness temperature would lead to an overestimation of
$\delta_\mathrm{var}$ in the $\gamma$-ray bright blazars. This alternative
explanation would not, however, explain the faster apparent jet speeds or
wider apparent jet opening angles of the LAT-detected sources.

The observed difference in the comoving-frame viewing angle distributions
between the $\gamma$-ray bright and weak blazars is an unanticipated
result. The left panel of Fig.~\ref{fig2} shows that the lack of LAT-detected
blazars at large comoving-frame viewing angles can be explained by low Doppler
factors of the sources at large $\theta_\mathrm{src}$. The beaming model does
not, however, explain the lack of LAT-detected sources at {\it small} values
of $\theta_\mathrm{src}$. If this lack is real, it may reflect an intrinsic
anisotropy of the $\gamma$-ray emission in the comoving frame of the jet. This
would have wide implications for the theoretical models of the high energy
emission from blazars, since almost all of these models rely on the assumption
that the $\gamma$-ray emission is (nearly) isotropic in the rest frame of the
relativistically moving sub-volume \citep[e.g.][]{der95}. Possible sources of
anisotropy in the $\gamma$-ray emission include, for example, anisotropic
$\gamma\gamma$ absorption or anisotropic seed photon field for inverse Compton
scattering.

The right hand panel of Fig.~\ref{fig2} shows that the non-LAT-detected
sources at small comoving frame viewing angles have small Lorentz factors,
except for \object{B0804+499}. This may provide an alternative explanation for
the apparent lack of $\gamma$-ray bright sources at small comoving frame
viewing angles if the \textit{intrinsic} $\gamma$-ray luminosity depends on
the bulk Lorentz factor in addition to being relativistically beamed. The fact
that \object{B0804+499} is not a bright $\gamma$-ray source, despite having a
very high Doppler factor and a moderately high Lorentz factor, poses a problem
for this explanation, but does not rule it out since the $\gamma$-ray emission
may be intermittent (or there may be something unusual in this particular
source).

\section{Summary}

We have investigated the connection between the $\gamma$-ray emission of
blazars and the intrinsic properties of their parsec-scale radio jets. Our
study, based on the 3-month \textit{Fermi} LAT bright gamma-ray source list,
shows that the $\gamma$-ray bright blazars have higher Doppler factors than
the $\gamma$-ray weak ones and confirms the earlier results from the EGRET era
\citep[e.g.,][]{lah03}.

It was also found that the distributions of the viewing angles in the comoving
frame of the jet differ significantly between the $\gamma$-ray bright and weak
blazars. While the lack of $\gamma$-ray-detected sources at large
comoving-frame viewing angles can be explained by $\gamma$-ray blazars being
more highly beamed, the apparent lack of $\gamma$-ray-detected sources at
small comoving-frame viewing angles instead hints of either an intrinsic
anisotropy or a Lorentz factor dependence of the $\gamma$-ray
emission. Unfortunately, our small sample size and the detection/non-detection
nature of the $\gamma$-ray data used in this analysis do not allow firm
conclusions to be drawn about the statistical significance of the lack of
$\gamma$-ray-detected sources at small comoving frame viewing angles if the
effects of beaming are taken into account. Since the potential intrinsic
emission anisotropy would have particularly important implications for the
theoretical models of the $\gamma$-ray production in blazars, further
investigation of a larger sample using more extensive $\gamma$-ray flux data
is clearly warranted and is being planned on the basis of \textit{Fermi}
1-year data.

\begin{acknowledgements}
  We thank Anne L\"ahteenm\"aki and the Mets\"ahovi monitoring project for
  providing unpublished data on two blazars. We thank Charles Dermer for
  discussions, as well as Ken Kellermann, Andrei Lobanov and Esko Valtaoja for
  commenting the manuscript. The MOJAVE project is supported under National
  Science Foundation grant AST-0807860 and NASA {\it Fermi} grant
  NNX08AV67G. T.~S.\ is a research fellow of the Alexander von~Humboldt
  Foundation. T.~S.\ also acknowledges a support by the Academy of Finland
  grant 120516. D.~C.~H.\ was supported by NSF grant AST-0707693.  Y.~Y.~K.\
  is partly supported by the Alexander von Humboldt return fellowship as well
  as by the Russian Foundation for Basic Research grant 08-02-00545. T.H.\
  acknowledges the support of the Academy of Finland for the Mets\"ahovi
  observing project. The VLBA is a facility of the National Science Foundation
  operated by the National Radio Astronomy Observatory under cooperative
  agreement with Associated Universities, Inc.
\end{acknowledgements}

\bibliographystyle{aa}
\bibliography{gamma}

\longtab{1}{
\begin{longtable}{lcccccccc}
\caption{Jet properties of the blazars in the Mets\"ahovi-MOJAVE sample \label{tab1}}\\
\hline \hline 
  IAU name & Class & $z$ & $\beta_\mathrm{app}^{(1)}$ & $\delta_\mathrm{var}$ &
  $\Gamma$ & $\theta$ &$\theta_\mathrm{src}$ &  LAT-det.$^{(2)}$ \\
  (B1950.0) & & & & & & (deg) & (deg) & \\ \hline
\endfirsthead
\caption{Continued.} \\
\hline
  IAU name & Class & $z$ & $\beta_\mathrm{app}^{(1)}$ & $\delta_\mathrm{var}$ &
  $\Gamma$ & $\theta$ &$\theta_\mathrm{src}$ &  LAT-det.$^{(2)}$ \\
  (B1950.0) & & &  & & & (deg) & (deg) &  \\ \hline
\endhead
\hline
\endfoot
\hline
\endlastfoot
  \object{0003$-$066} & BL Lac &  0.347 &   0.9  &   5.1  &   2.7 &    4.0 &   20.5 &   N \\
  \object{0016+731} & Quasar &  1.781 &   6.7  &   7.8  &   6.9 &    7.3 &   82.0 &   N \\
  \object{0106+013} & Quasar &  2.099 &  26.5  &  18.2  &  28.4 &    2.9 &  111.0 &   N \\
  \object{0109+224} & BL Lac &  0.265 &   ...  &   9.1  &   ... &    ... &    ... &   Y \\
  \object{0133+476} & Quasar &  0.859 &  13.0  &  20.5  &  14.4 &    2.5 &   64.8 &   Y \\
  \object{0202+149} & Quasar &  0.405 &   6.4  &  15.0  &   8.9 &    2.8 &   46.6 &   N \\
  \object{0212+735} & Quasar &  2.367 &   7.6  &   8.4  &   7.7 &    6.8 &   84.9 &   N \\
  \object{0234+285} & Quasar &  1.207 &  12.3  &  16.0  &  12.7 &    3.5 &   75.4 &   Y \\
  \object{0235+164} & BL Lac &  0.940 &   ...  &  23.8  &   ... &    ... &    ... &   Y \\
  \object{0333+321} & Quasar &  1.259 &  12.8  &  22.0  &  14.7 &    2.3 &   60.5 &   N \\
  \object{0336$-$019} & Quasar &  0.852 &  22.4  &  17.2  &  23.2 &    3.2 &  104.9 &   N \\
  \object{0420$-$014} & Quasar &  0.915 &   7.3  &  19.7  &  11.3 &    1.9 &   41.0 &   Y \\
  \object{0458$-$020} & Quasar &  2.286 &  16.5  &  15.7  &  16.6 &    3.7 &   93.1 &   N \\
  \object{0528+134} & Quasar &  2.070 &  19.2  &  30.9  &  21.4 &    1.7 &   63.7 &   Y \\
  \object{0605$-$085} & Quasar &  0.872 &  16.8  &   7.5  &  22.6 &    5.7 &  131.8 &   N \\
  \object{0642+449} & Quasar &  3.396 &   0.8  &  10.6  &   5.4 &    0.8 &    8.3 &   N \\
  \object{0716+714} & BL Lac &  0.310 &  10.1  &  10.8  &  10.2 &    5.3 &   86.4 &   Y \\
  \object{0736+017} & Quasar &  0.191 &  14.4  &   8.5  &  16.5 &    5.9 &  119.0 &   N \\
  \object{0754+100} & BL Lac &  0.266 &  14.4  &   5.5  &  21.6 &    6.9 &  138.0 &   N \\
  \object{0804+499} & Quasar &  1.436 &   1.8  &  35.2  &  17.6 &    0.2 &    6.0 &   N \\
  \object{0814+425} & BL Lac &  0.245 &   1.7  &   4.6  &   2.7 &    8.6 &   42.8 &   Y \\
  \object{0827+243} & Quasar &  0.941 &  22.0  &  13.0  &  25.2 &    3.9 &  119.0 &   N \\
  \object{0836+710} & Quasar &  2.218 &  25.4  &  16.1  &  28.1 &    3.2 &  115.2 &   N \\
  \object{0851+202} & BL Lac &  0.306 &   5.2  &  16.8  &   9.3 &    1.9 &   34.5 &   Y \\
  \object{0923+392} & Quasar &  0.695 &   0.6  &   4.3  &   2.3 &    3.9 &   16.7 &   N \\
  \object{0945+408} & Quasar &  1.249 &  18.6  &   6.3  &  30.5 &    5.5 &  142.4 &   N \\
  \object{1055+018} & Quasar &  0.888 &   8.1  &  12.1  &   8.8 &    4.4 &   67.6 &   Y \\
  \object{1156+295} & Quasar &  0.729 &  24.9  &  28.2  &  25.1 &    2.0 &   82.9 &   Y \\
  \object{1222+216} & Quasar &  0.432 &  21.0  &   5.2  &  45.5 &    5.1 &  152.5 &   N \\
  \object{1226+023} & Quasar &  0.158 &  13.4  &  16.8  &  13.8 &    3.3 &   77.1 &   Y \\
  \object{1253$-$055} & Quasar &  0.536 &  20.6  &  23.8  &  20.8 &    2.4 &   81.9 &   Y \\
  \object{1308+326} & Quasar &  0.997 &  20.9  &  15.3  &  22.0 &    3.6 &  107.8 &   Y \\
  \object{1324+224} & Quasar &  1.400 &   ...  &  21.0  &   ... &    ... &    ... &   N \\
  \object{1334$-$127} & Quasar &  0.539 &  10.3  &   8.3  &  10.6 &    6.8 &  102.6 &   N \\ 
  \object{1413+135} & BL Lac &  0.247 &   1.8  &  12.1  &   6.2 &    1.4 &   17.1 &   N \\
  \object{1502+106} & Quasar &  1.839 &  14.8  &  11.9  &  15.2 &    4.7 &  102.6 &   Y \\
  \object{1510$-$089} & Quasar &  0.360 &  20.2  &  16.5  &  20.6 &    3.4 &  101.4 &   Y \\
  \object{1538+149} & BL Lac &  0.605 &   8.7  &   4.3  &  11.2 &   10.6 &  128.5 &   N \\
  \object{1606+106} & Quasar &  1.226 &  17.9  &  24.8  &  18.9 &    2.2 &   71.8 &   N \\
  \object{1611+343} & Quasar &  1.397 &   5.7  &  13.6  &   8.0 &    3.0 &   45.9 &   N \\
  \object{1633+382} & Quasar &  1.814 &  29.5  &  21.3  &  31.1 &    2.6 &  108.4 &   Y \\
  \object{1637+574} & Quasar &  0.751 &  10.6  &  13.9  &  11.0 &    4.0 &   75.0 &   N \\
  \object{1641+399} & Quasar &  0.593 &  19.3  &   7.7  &  28.0 &    5.1 &  136.4 &   N \\
  \object{1730$-$130} & Quasar &  0.902 &  35.7  &  10.6  &  65.5 &    2.9 &  146.9 &   N \\
  \object{1739+522} & Quasar &  1.379 &   ...  &  26.3  &   ... &    ... &    ... &   N \\
  \object{1741$-$038} & Quasar &  1.054 &   ...  &  19.5  &   ... &    ... &    ... &   N \\
  \object{1749+096} & BL Lac &  0.322 &   6.8  &  11.9  &   8.0 &    4.2 &   60.1 &   Y \\
  \object{1803+784} & BL Lac &  0.684 &   9.0  &  12.1  &   9.4 &    4.5 &   73.4 &   Y \\  
  \object{1807+698} & BL Lac &  0.051 &   0.1  &   1.1  &   1.0 &   45.3 &   50.8 &   N \\
  \object{1823+568} & BL Lac &  0.664 &   9.4  &   6.3  &  10.2 &    8.4 &  112.4 &   N \\
  \object{1828+487} & Quasar &  0.692 &  13.7  &   5.6  &  19.5 &    7.1 &  135.4 &   N \\
  \object{1928+738} & Quasar &  0.302 &   7.2  &   1.9  &  15.0 &   14.8 &  151.2 &   N \\
  \object{2121+053} & Quasar &  1.941 &   8.4  &  15.2  &   9.9 &    3.2 &   58.1 &   N \\
  \object{2134+004} & Quasar &  1.932 &   2.0  &  16.0  &   8.1 &    0.9 &   14.6 &   N \\
  \object{2136+141} & Quasar &  2.427 &   3.0  &   8.2  &   4.7 &    4.5 &   39.9 &   N \\
  \object{2145+067} & Quasar &  0.990 &   2.2  &  15.5  &   7.9 &    1.0 &   16.2 &   N \\
  \object{2200+420} & BL Lac &  0.069 &   5.0  &   7.2  &   5.4 &    7.5 &   69.7 &   Y \\
  \object{2201+315} & Quasar &  0.295 &   7.9  &   6.6  &   8.1 &    8.5 &  100.3 &   N \\
  \object{2223$-$052} & Quasar &  1.404 &  14.6  &  15.9  &  14.7 &    3.6 &   85.4 &   N \\
  \object{2227$-$088} & Quasar &  1.562 &   8.1  &  15.8  &  10.0 &    3.0 &   54.8 &   Y \\
  \object{2230+114} & Quasar &  1.037 &  15.4  &  15.5  &  15.4 &    3.7 &   89.9 &   Y \\
  \object{2251+158} & Quasar &  0.859 &  14.2  &  32.9  &  19.5 &    1.3 &   46.7 &   Y \\
\end{longtable} {\footnotesize \noindent $^{(1)}$ Fastest non-accelerating,
  radial apparent speed measured in the source. The speeds differ in some
  cases from the speeds quoted by \citet{hov09}, partly because Hovatta et
  al. used results from a preliminary kinematics analysis, and partly because
  we consider only non-accelerating, radially moving components in order to
  estimate the flow speed close to the core, where the mm-wavelength outbursts
  take place. Due to this, the values of $\Gamma$ and $\theta$ naturally also
  differ in some cases from those reported by Hovatta et al.\\
  $^{(2)}$ Bright ($>10\sigma$) LAT-detections during the first three months
  of science operations \citep{abd09a}.  }}

\end{document}